\documentclass[12pt]{article}
\usepackage{feynarts}
\usepackage{a4wide,epsfig}
\newcommand{\agt}{\,\rlap{\lower 3.5 pt \hbox{$\mathchar \sim$}} \raise 1pt
 \hbox {$>$}\,}
\newcommand{\alt}{\,\rlap{\lower 3.5 pt \hbox{$\mathchar \sim$}} \raise 1pt
 \hbox {$<$}\,}

\begin{document}

\title{
\vskip-3cm{\baselineskip14pt
\centerline{\normalsize DESY 04-040\hfill ISSN 0418-9833}
\centerline{\normalsize hep-ph/0405232\hfill}
\centerline{\normalsize May 2004\hfill}}
\vskip1.5cm
Two-loop electroweak correction of $\mathcal{O}(G_{F} M_{t}^{2})$ to the
Higgs-boson decay into photons}

\author{Frank Fugel, Bernd A. Kniehl, Matthias Steinhauser
  \\[2em]
{\normalsize II. Institut f\"ur Theoretische Physik, Universit\"at Hamburg,}\\
{\normalsize Luruper Chaussee 149, 22761 Hamburg, Germany}
}
\date{}
\maketitle

\begin{abstract}
We compute the dominant two-loop electroweak correction, of
$\mathcal{O}(G_{F} M_{t}^{2})$, to the partial width of the decay of an
intermediate-mass Higgs boson into a pair of photons.
We use the asymptotic-expansion technique in order to extract the leading
dependence on the top-quark mass plus four expansion terms that describe the
dependence on the $W$- and Higgs-boson masses.
This correction reduces the Born result by approximately 2.5\%.
As a by-product of our analysis, we also recover the
$\mathcal{O}(G_{F} M_{t}^{2})$ correction to the partial width of the
Higgs-boson decay to two gluon jets.

\medskip

\end{abstract}

\newpage


\section{\label{sec::intro}Introduction}

Among the main tasks of the current experiments at the Fermilab Tevatron and
the future experiments at the CERN Large Hadron Collider (LHC) is the search
for the Higgs boson, which is the only missing particle in the standard model
(SM).
The electroweak precision data mainly collected at CERN LEP and SLAC SLC in
combination with the direct top-quark mass measurement at the Tevatron favour
a light Higgs boson with mass $M_H=113^{+62}_{-42}$~GeV with an upper bound of
about 237~GeV at the 95\% confidence level~\cite{ewwg}.
The allowed mass range is compatible with the so-called intermediate mass
range, defined by $M_W\le M_H\le 2M_W$.
In this mass range, the decay into two photons represents one of the most
useful detection modes at hadron colliders.

\begin{figure}[t]
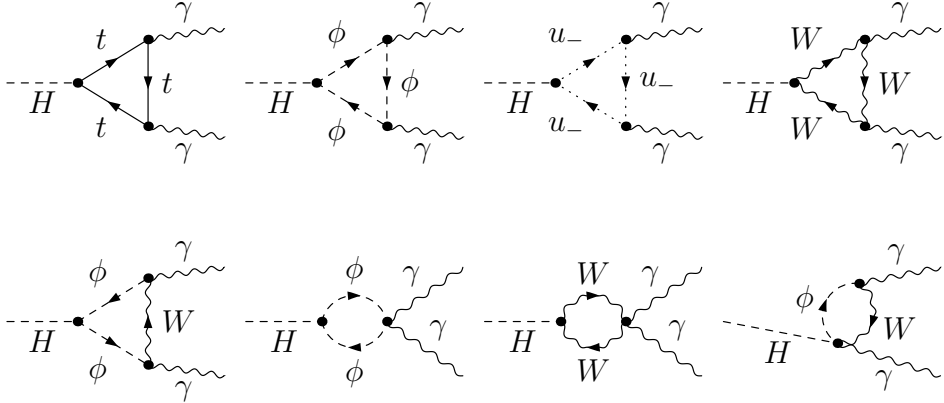

\begin{center}

\unitlength=1bp%

\begin{feynartspicture}(370,180)(4,2)

\FADiagram{}
\FAProp(0.,10.)(6.5,10.)(0.,){/ScalarDash}{0}
\FALabel(3.25,9.18)[t]{$H$}
\FAProp(20.,15.)(13.,14.)(0.,){/Sine}{0}
\FALabel(16.2808,15.5544)[b]{$\gamma$}
\FAProp(20.,5.)(13.,6.)(0.,){/Sine}{0}
\FALabel(16.2808,4.44558)[t]{$\gamma$}
\FAProp(6.5,10.)(13.,14.)(0.,){/Straight}{1}
\FALabel(9.20801,13.1807)[br]{$t$}
\FAProp(6.5,10.)(13.,6.)(0.,){/Straight}{-1}
\FALabel(9.20801,6.81927)[tr]{$t$}
\FAProp(13.,14.)(13.,6.)(0.,){/Straight}{1}
\FALabel(14.274,10.)[l]{$t$}
\FAVert(6.5,10.){0}
\FAVert(13.,14.){0}
\FAVert(13.,6.){0}

\FADiagram{}
\FAProp(0.,10.)(6.5,10.)(0.,){/ScalarDash}{0}
\FALabel(3.25,9.18)[t]{$H$}
\FAProp(20.,15.)(13.,14.)(0.,){/Sine}{0}
\FALabel(16.2808,15.5544)[b]{$\gamma$}
\FAProp(20.,5.)(13.,6.)(0.,){/Sine}{0}
\FALabel(16.2808,4.44558)[t]{$\gamma$}
\FAProp(6.5,10.)(13.,14.)(0.,){/ScalarDash}{1}
\FALabel(9.20801,13.1807)[br]{$\phi$}
\FAProp(6.5,10.)(13.,6.)(0.,){/ScalarDash}{-1}
\FALabel(9.20801,6.81927)[tr]{$\phi$}
\FAProp(13.,14.)(13.,6.)(0.,){/ScalarDash}{1}
\FALabel(14.274,10.)[l]{$\phi$}
\FAVert(6.5,10.){0}
\FAVert(13.,14.){0}
\FAVert(13.,6.){0}

\FADiagram{}
\FAProp(0.,10.)(6.5,10.)(0.,){/ScalarDash}{0}
\FALabel(3.25,9.18)[t]{$H$}
\FAProp(20.,15.)(13.,14.)(0.,){/Sine}{0}
\FALabel(16.2808,15.5544)[b]{$\gamma$}
\FAProp(20.,5.)(13.,6.)(0.,){/Sine}{0}
\FALabel(16.2808,4.44558)[t]{$\gamma$}
\FAProp(6.5,10.)(13.,14.)(0.,){/GhostDash}{1}
\FALabel(9.20801,13.1807)[br]{$u_-$}
\FAProp(6.5,10.)(13.,6.)(0.,){/GhostDash}{-1}
\FALabel(9.20801,6.81927)[tr]{$u_-$}
\FAProp(13.,14.)(13.,6.)(0.,){/GhostDash}{1}
\FALabel(14.274,10.)[l]{$u_-$}
\FAVert(6.5,10.){0}
\FAVert(13.,14.){0}
\FAVert(13.,6.){0}

\FADiagram{}
\FAProp(0.,10.)(6.5,10.)(0.,){/ScalarDash}{0}
\FALabel(3.25,9.18)[t]{$H$}
\FAProp(20.,15.)(13.,14.)(0.,){/Sine}{0}
\FALabel(16.2808,15.5544)[b]{$\gamma$}
\FAProp(20.,5.)(13.,6.)(0.,){/Sine}{0}
\FALabel(16.2808,4.44558)[t]{$\gamma$}
\FAProp(6.5,10.)(13.,14.)(0.,){/Sine}{1}
\FALabel(9.20801,13.1807)[br]{$W$}
\FAProp(6.5,10.)(13.,6.)(0.,){/Sine}{-1}
\FALabel(9.20801,6.81927)[tr]{$W$}
\FAProp(13.,14.)(13.,6.)(0.,){/Sine}{1}
\FALabel(14.274,10.)[l]{$W$}
\FAVert(6.5,10.){0}
\FAVert(13.,14.){0}
\FAVert(13.,6.){0}

\FADiagram{}
\FAProp(0.,10.)(6.5,10.)(0.,){/ScalarDash}{0}
\FALabel(3.25,9.18)[t]{$H$}
\FAProp(20.,15.)(13.,14.)(0.,){/Sine}{0}
\FALabel(16.2808,15.5544)[b]{$\gamma$}
\FAProp(20.,5.)(13.,6.)(0.,){/Sine}{0}
\FALabel(16.2808,4.44558)[t]{$\gamma$}
\FAProp(6.5,10.)(13.,14.)(0.,){/ScalarDash}{-1}
\FALabel(9.20801,13.1807)[br]{$\phi$}
\FAProp(6.5,10.)(13.,6.)(0.,){/ScalarDash}{1}
\FALabel(9.20801,6.81927)[tr]{$\phi$}
\FAProp(13.,14.)(13.,6.)(0.,){/Sine}{-1}
\FALabel(14.274,10.)[l]{$W$}
\FAVert(6.5,10.){0}
\FAVert(13.,14.){0}
\FAVert(13.,6.){0}

\FADiagram{}
\FAProp(0.,10.)(7.,10.)(0.,){/ScalarDash}{0}
\FALabel(3.5,9.18)[t]{$H$}
\FAProp(20.,15.)(13.,10.)(0.,){/Sine}{0}
\FALabel(16.0791,13.2813)[br]{$\gamma$}
\FAProp(20.,5.)(13.,10.)(0.,){/Sine}{0}
\FALabel(16.9209,8.28129)[bl]{$\gamma$}
\FAProp(7.,10.)(13.,10.)(0.8,){/ScalarDash}{-1}
\FALabel(10.,6.53)[t]{$\phi$}
\FAProp(7.,10.)(13.,10.)(-0.8,){/ScalarDash}{1}
\FALabel(10.,13.47)[b]{$\phi$}
\FAVert(7.,10.){0}
\FAVert(13.,10.){0}

\FADiagram{}
\FAProp(0.,10.)(7.,10.)(0.,){/ScalarDash}{0}
\FALabel(3.5,9.18)[t]{$H$}
\FAProp(20.,15.)(13.,10.)(0.,){/Sine}{0}
\FALabel(16.0791,13.2813)[br]{$\gamma$}
\FAProp(20.,5.)(13.,10.)(0.,){/Sine}{0}
\FALabel(16.9209,8.28129)[bl]{$\gamma$}
\FAProp(7.,10.)(13.,10.)(0.8,){/Sine}{-1}
\FALabel(10.,6.53)[t]{$W$}
\FAProp(7.,10.)(13.,10.)(-0.8,){/Sine}{1}
\FALabel(10.,13.47)[b]{$W$}
\FAVert(7.,10.){0}
\FAVert(13.,10.){0}

\FADiagram{}
\FAProp(0.,10.)(10.5,8.)(0.,){/ScalarDash}{0}
\FALabel(5.00675,8.20296)[t]{$H$}
\FAProp(20.,15.)(12.5,13.5)(0.,){/Sine}{0}
\FALabel(15.946,15.2899)[b]{$\gamma$}
\FAProp(20.,5.)(10.5,8.)(0.,){/Sine}{0}
\FALabel(14.7832,5.50195)[t]{$\gamma$}
\FAProp(12.5,13.5)(10.5,8.)(0.8,){/ScalarDash}{-1}
\FALabel(8.32332,12.0797)[r]{$\phi$}
\FAProp(12.5,13.5)(10.5,8.)(-0.8,){/Sine}{1}
\FALabel(14.6767,9.4203)[l]{$W$}
\FAVert(12.5,13.5){0}
\FAVert(10.5,8.){0}

\end{feynartspicture}

\caption{\label{1loop} Sample Feynman diagrams contributing in leading order
to the process $H\to\gamma\gamma$.}

\end{center}
\end{figure}

Since there is no direct coupling of the Higgs boson to photons, the process
$H\to\gamma\gamma$ is loop-induced.
In the limit of vanishing bottom-quark mass, one distinguishes at lowest order
the contributions from virtual top quarks and $W$ bosons, where, in covariant
gauge, the latter are accompanied by charged Goldstone bosons ($\phi$) and
Faddeev-Popov ghosts ($u$).
Some sample Feynman diagrams are depicted in Fig.~\ref{1loop}. 
The corresponding contributions have been evaluated for the first time in
Ref.~\cite{Ellis:1975ap} (for reviews, see also Ref.~\cite{Kniehl:1993ay}).
QCD corrections, which only affect the top-quark diagrams are known at the
two-~\cite{Djouadi:1990aj} and three-loop~\cite{Ste96} orders.
Recently, also the two-loop electroweak correction induced by light-fermion
loops has been evaluated~\cite{Aglietti:2004nj}.
In this paper, we compute the two-loop electroweak correction that is
enhanced by $G_F M_t^2$.
For this purpose, we consider the (formal) hierarchy
$M_t^2\gg (2M_W)^2 \gg M_H^2$ and apply the method of asymptotic
expansion~\cite{Smirnov:pj}, which allows us to also evaluate four expansion
terms in the ratio $\tau_W=M_H^2/(2M_W)^2$ aside from the leading term in
$M_t^2$.

Due to electromagnetic gauge invariance, the amputated transition-matrix
element of $H \to \gamma \gamma$ possesses the structure
\begin{eqnarray}
  \mathcal{T}^{\mu \nu} &=&
  (q_{1}\!\cdot\!q_{2} \, g^{\mu \nu} - q_{1}^{\nu}q_{2}^{\mu}) \mathcal{A},
  \label{Amplitude}
\end{eqnarray}
where $\mu$ and $\nu$ are the Lorentz indices of the external photons with
four-momenta $q_{1}$ and $q_{2}$, respectively.
Thus, the decay rate of the Higgs boson into two photons is given by
\begin{equation}
  \Gamma(H \to \gamma \gamma) = \frac{M_{H}^{3}}{64 \pi}|\mathcal{A}|^2.
\end{equation}
The form factor $\mathcal{A}$ is evaluated in perturbation theory as
\begin{equation}
\mathcal{A}=\mathcal{A}_{t}^{(0)} 
  + \mathcal{A}_{W}^{(0)} 
  + \mathcal{A}_{tW}^{(1)} + \cdots,
  \label{Notation}
\end{equation}
where ${\cal A}_t^{(0)}$ and ${\cal A}_W^{(0)}$ denote the one-loop 
contributions induced by virtual top quarks and $W$ bosons, respectively,
${\cal A}_{tW}^{(1)}$ stands for the two-loop electroweak correction involving
virtual top quarks, and the ellipsis represents the residual one- and two-loop
contributions as well as all contributions involving more than two loops.

In the practical calculation, it is convenient to project out the scalar
amplitudes that multiply the basic Lorentz tensors $g^{\mu \nu}$,
$q_1^{\mu} q_2^{\nu}$, and $q_1^{\nu} q_2^{\mu}$.
The corresponding projectors can be obtained by in turn contracting
$\mathcal{T}^{\mu\nu}$ with these Lorentz tensors and solving the resulting
system of linear equations.
We separately project out the coefficients of the tensors
$q_{1}\!\cdot\!q_{2}\,g^{\mu \nu}$ and $q_{1}^{\nu}q_{2}^{\mu}$, and thus have
a strong check on our calculation.
Furthermore, we adopt a general $R_{\xi}$ gauge in our calculation and verify
that the gauge parameter drops out in the final result.
For simplicity, the element $V_{tb}$ of the Cabibbo-Kobayashi-Maskawa quark
mixing matrix is set to unity, so that the quarks of the third fermion
generation decouple from those of the first two, which they actually do to
very good approximation \cite{Hagiwara:fs}.


Our paper is organized as follows.
In Section~\ref{sec::1loop}, we illustrate the usefulness of the
asymptotic-expansion technique by redoing the one-loop calculation.
In Section~\ref{sec::2loop}, we discuss the two-loop calculation.
Section~\ref{sec::numerics} contains the discussion of the numerical results.
We conclude with a summary in Section~\ref{sec::summary}.


\section{\label{sec::1loop}One-loop results and counterterm contribution}

The application of the asymptotic-expansion technique to the one-loop
diagrams, some of which are shown in Fig.~\ref{1loop}, leads to a naive Taylor
expansion in the external momenta $q_1$ and $q_2$.
Nevertheless, we use already here a completely automated set-up, which
consists in the successive use of the computer programs
\texttt{QGRAF}~\cite{Nogueira:1991ex}, \texttt{q2e}~\cite{Seidensticker},
\texttt{exp}~\cite{Harlander:1997zb}, and
\texttt{MATAD}~\cite{Steinhauser:2000ry}.
First, \texttt{QGRAF} is used to generate the Feynman diagrams.
Its output is then rewritten by \texttt{q2e} to be understandable by
\texttt{exp}.
In the two-loop case, the latter performs the asymptotic expansion and
generates the relevant subdiagrams according to the rules of the so-called
hard-mass procedure~\cite{Smirnov:pj}.
\texttt{Form}~\cite{Vermaseren} files are generated, which can be read by
\texttt{MATAD}~\cite{Steinhauser:2000ry}, which performs the very calculation
of the diagrams.

The analytic expression for the Born result can be found in
Refs.~\cite{Ellis:1975ap,Shi79}.
For completeness, we list it here in closed form and as an expansion for
$(2M_W)^2\gg M_H^2$, which we reproduce.
One has
\begin{eqnarray}
  \mathcal{A}_{t}^{(0)} & = & 
  \hat{\mathcal{A}} N_c Q_t^2 \left\{ \frac{1}{\tau_{t}} \left[ 1 +
  \left( 1 - \frac{1}{\tau_{t}} \right) \arcsin^{2}\sqrt{\tau_{t}}
  \right] \right\}
  \nonumber\\ 
  & = & \hat{\mathcal{A}} N_c Q_t^2 \left( \frac{2}{3} + \frac{7}{45} \tau_{t}
  + \frac{4}{63} \tau_{t}^{2} 
  + \frac{52}{1575} \tau_{t}^{3} 
  + \frac{1024}{51975} \tau_{t}^{4} 
  + \frac{2432}{189189} \tau_{t}^{5} 
  + \ldots \right),
  \\ 
  \mathcal{A}_{W}^{(0)} & = & 
  \hat{\mathcal{A}} \left\{ - \frac{1}{2} \left[ 2 +
  \frac{3}{\tau_{W}} + \frac{3}{\tau_{W}} \left( 2 -
  \frac{1}{\tau_{W}} \right) \arcsin^{2}\sqrt{\tau_{W}} \right]
  \right\}
  \nonumber\\
  & = & \hat{\mathcal{A}} \left( - \frac{7}{2} -
  \frac{11}{15} \tau_{W} 
  - \frac{38}{105} \tau_{W}^{2} 
  - \frac{116}{525} \tau_{W}^{3} 
  - \frac{2624}{17325} \tau_{W}^{4} 
  - \frac{640}{5733} \tau_{W}^{5} 
  + \ldots \right),
\label{eq::born}
\end{eqnarray}
where $\hat{\mathcal{A}} = 2^{1/4} G_{F}^{1/2}(\alpha/\pi)$,
$\tau_{t} = M_{H}^{2}/(2 M_{t})^{2}$, and $\tau_{W}$ is defined in
Section~\ref{sec::intro}.
Here, $\alpha$ is Sommerfeld's fine-structure constant, $G_F$ is Fermi's
constant, $N_{c}=3$ is the number of quark colours, and $Q_{t}=2/3$ is the
electric charge of the top quark in units of the positron charge.
The Higgs-boson mass entering the expansion parameters $\tau_{t}$ and
$\tau_{W}$ partly arises from the couplings involving one Higgs and two
Goldstone bosons, but also from the expansion in the external momenta due to
the kinematic relation $(q_{1} + q_{2})^{2} = M_{H}^{2}$. 

We wish to note that the result for the process $H\to gg$ is simply obtained
by setting $\mathcal{A}_W^{(0)}=0$ and performing the substitution 
$\alpha N_{c} Q_{t}^{2} \to \alpha_{s} \sqrt{2}$, where $\alpha_s$ is the
strong-coupling constant.

In the remainder of this section, let us discuss the counterterm contributions
needed for our two-loop analysis.
We adopt the on-mass-shell scheme and regularize the ultraviolet divergences by
means of dimensional regularization, with $D=4-2 \epsilon$ space-time
dimensions and 't~Hooft mass scale $\mu$.
We use the anti-commuting definition of $\gamma_5$.

As will become clearer in the next section, it is important to treat the
tadpole contributions properly in our calculation.
For this reason, in the following, we list the corresponding contributions
separately and mark them by the superscript ``tad''.
Inspection of the one-loop diagrams reveals that, to $\mathcal{O}(G_F M_t^2)$,
we have to renormalize the Higgs-boson wave function and mass, the $W$-boson
mass, and the top-quark mass.
The corresponding counterterms are defined through
\begin{eqnarray}
  H^0 &=& \sqrt{Z_H} H = \left(1+\frac{1}{2} \delta Z_H\right) H ,
  \nonumber \\
  (M_H^0)^2 &=& M_H^2 + \delta M_H^2 + \delta M_H^{2,{\rm tad}} ,
  \nonumber \\
  (M_W^0)^2 &=& M_W^2 + \delta M_W^2 + \delta M_W^{2,{\rm tad}} ,
  \nonumber \\
  M_t^0 &=& M_t + \delta M_t + \delta M_t^{\rm tad} .
\end{eqnarray}
Note that $\delta Z_H$ is obtained from the derivative of the Higgs-boson
self-energy and thus has no tadpole contribution.
The mass counterterms are obtained from the corresponding two-point functions,
where only the $M_t$-enhanced contributions have to be considered.
Up to and including quadratic terms in $M_t$, we have
\begin{eqnarray}
  \delta Z_H &=& -2 N_c x_t \left( \Delta - \ln\frac{M_{t}^{2}}{\mu^{2}} -
  \frac{2}{3} \right),
  \nonumber\\
  \frac{\delta M_{H}^{2}}{M_{H}^{2}} & = & N_c x_t
  \left[ -12 \frac{M_t^2}{M_{H}^{2}} \left( \Delta -
  \ln\frac{M_{t}^{2}}{\mu^{2}}+ \frac{1}{3} \right) 
    + 2 \left( \Delta - \ln \frac{M_{t}^{2}}{\mu^{2}} 
    - \frac{2}{3} \right) \right],
\nonumber\\
  \frac{\delta M_{H}^{2,tad}}{M_{H}^{2}} & = & 12 N_{c} x_t
  \frac{M_t^2}{M_{H}^{2}}
  \left( \Delta - \ln\frac{M_{t}^{2}}{\mu^{2}} + 1\right) ,
\nonumber\\
  \frac{\delta M_{W}^{2}}{M_{W}^{2}} &=& -2 N_c x_t
  \left( \Delta - \ln\frac{M_{t}^{2}}{\mu^{2}} + \frac{1}{2} \right) ,
\nonumber\\
  \frac{\delta M_{W}^{2,tad}}{M_{W}^{2}} &=& 8 N_c x_t
  \frac{M_{t}^{2}}{M_{H}^{2}}
  \left( \Delta - \ln\frac{M_{t}^{2}}{\mu^{2}}  + 1 \right) ,
\nonumber\\
  \frac{\delta M_{t}}{M_{t}} &=& \frac{3}{2} x_t 
  \left( \Delta - \ln\frac{M_{t}^{2}}{\mu^{2}} 
  + \frac{8}{3} \right) ,
  \nonumber\\
  \frac{\delta M_{t}^{tad}}{M_{t}} &=& 4 N_c x_t \frac{M_{t}^{2}}{M_{H}^{2}}
  \left( \Delta - \ln\frac{M_{t}^{2}}{\mu^{2}}+ 1 \right) ,
\label{eq:ct}
\end{eqnarray}
with $x_t=G_FM_t^2/(8\pi^2\sqrt{2})$ and
$\Delta = 1/\epsilon - \gamma_{E} + \ln (4 \pi)$,
where $\gamma_{E}$ is Euler's constant.
The $W$-boson mass renormalization is needed for the $W$, $\phi$, and $u$
propagators, where $M_W$ enters as a parameter.
Furthermore, also the $HW^{\pm}W^{\mp}$, $H\phi^{\pm}W^{\mp}$,
$\phi^{\pm}W^{\mp}\gamma$, and $H\phi^{\pm}W^{\mp}\gamma$ vertices contain
$M_W$.
The only vertex involving $M_H$ is $H\phi^{\pm}\phi^{\mp}$, which induces
two-loop contributions via $\delta M_H$.
Finally, $M_t$ occurs in the top-quark propagator and in the $Ht\overline{t}$
vertex.

The wave function renormalization and the renormalization of a factor
$1/M_W$ is common to all one-loop diagrams.
This allows for the definition of a universal factor \cite{hll}, which is
finite in our calculation.
It is given by
\begin{eqnarray}
  \delta_u &=& \frac{1}{2}\left(\delta Z_H - \frac{\delta
  M_W^2}{M_W^2} \right)
\nonumber\\
 & = & \frac{7}{6} N_c x_t .
  \label{RenUniv}
\end{eqnarray}


\section{\label{sec::2loop}Two-loop results}

\begin{figure}[th]
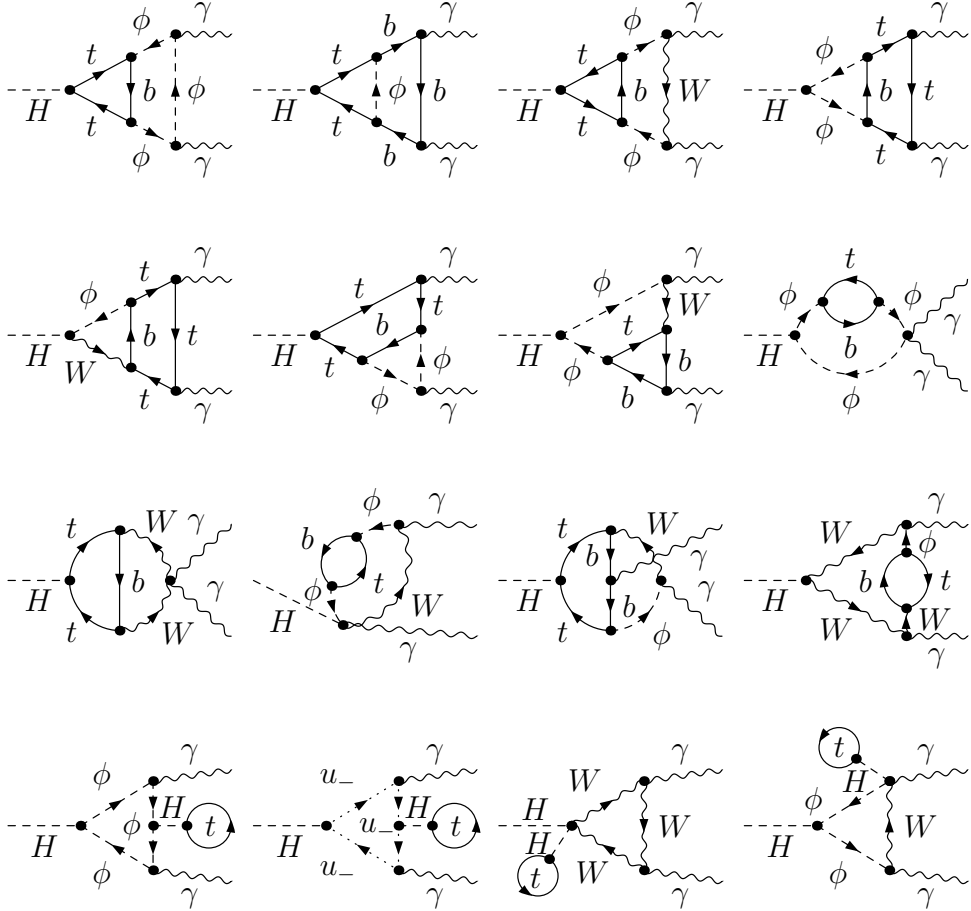

\begin{center}

\unitlength=1bp%

\begin{feynartspicture}(370,400)(4,4)

\FADiagram{}
\FAProp(0.,10.)(5.5,10.)(0.,){/ScalarDash}{0}
\FALabel(2.75,9.18)[t]{$H$}
\FAProp(20.,15.)(15.,15.)(0.,){/Sine}{0}
\FALabel(17.5,16.07)[b]{$\gamma$}
\FAProp(20.,5.)(15.,5.)(0.,){/Sine}{0}
\FALabel(17.5,3.93)[t]{$\gamma$}
\FAProp(11.,7.1)(5.5,10.)(0.,){/Straight}{1}
\FALabel(8.00707,7.65892)[tr]{$t$}
\FAProp(11.,7.1)(15.,5.)(0.,){/ScalarDash}{1}
\FALabel(12.7595,5.15763)[tr]{$\phi$}
\FAProp(11.,12.95)(11.,7.1)(0.,){/Straight}{1}
\FALabel(12.07,10.025)[l]{$b$}
\FAProp(11.,12.95)(5.5,10.)(0.,){/Straight}{-1}
\FALabel(7.99737,12.3609)[br]{$t$}
\FAProp(11.,12.95)(15.,15.)(0.,){/ScalarDash}{-1}
\FALabel(12.7731,14.8744)[br]{$\phi$}
\FAProp(15.,15.)(15.,5.)(0.,){/ScalarDash}{-1}
\FALabel(16.07,10.)[l]{$\phi$}
\FAVert(11.,12.95){0}
\FAVert(11.,7.1){0}
\FAVert(5.5,10.){0}
\FAVert(15.,15.){0}
\FAVert(15.,5.){0}

\FADiagram{}
\FAProp(0.,10.)(5.5,10.)(0.,){/ScalarDash}{0}
\FALabel(2.75,9.18)[t]{$H$}
\FAProp(20.,15.)(15.,15.)(0.,){/Sine}{0}
\FALabel(17.5,16.07)[b]{$\gamma$}
\FAProp(20.,5.)(15.,5.)(0.,){/Sine}{0}
\FALabel(17.5,3.93)[t]{$\gamma$}
\FAProp(11.,7.1)(5.5,10.)(0.,){/Straight}{1}
\FALabel(8.00707,7.65892)[tr]{$t$}
\FAProp(11.,7.1)(15.,5.)(0.,){/Straight}{-1}
\FALabel(12.7595,5.15763)[tr]{$b$}
\FAProp(11.,12.95)(11.,7.1)(0.,){/ScalarDash}{-1}
\FALabel(12.07,10.025)[l]{$\phi$}
\FAProp(11.,12.95)(5.5,10.)(0.,){/Straight}{-1}
\FALabel(7.99737,12.3609)[br]{$t$}
\FAProp(11.,12.95)(15.,15.)(0.,){/Straight}{1}
\FALabel(12.7731,14.8744)[br]{$b$}
\FAProp(15.,15.)(15.,5.)(0.,){/Straight}{1}
\FALabel(16.07,10.)[l]{$b$}
\FAVert(11.,12.95){0}
\FAVert(11.,7.1){0}
\FAVert(5.5,10.){0}
\FAVert(15.,15.){0}
\FAVert(15.,5.){0}

\FADiagram{}
\FAProp(0.,10.)(5.5,10.)(0.,){/ScalarDash}{0}
\FALabel(2.75,9.18)[t]{$H$}
\FAProp(20.,15.)(15.,15.)(0.,){/Sine}{0}
\FALabel(17.5,16.07)[b]{$\gamma$}
\FAProp(20.,5.)(15.,5.)(0.,){/Sine}{0}
\FALabel(17.5,3.93)[t]{$\gamma$}
\FAProp(11.,7.1)(5.5,10.)(0.,){/Straight}{-1}
\FALabel(8.00707,7.65892)[tr]{$t$}
\FAProp(11.,7.1)(15.,5.)(0.,){/ScalarDash}{-1}
\FALabel(12.7595,5.15763)[tr]{$\phi$}
\FAProp(11.,12.95)(11.,7.1)(0.,){/Straight}{-1}
\FALabel(12.07,10.025)[l]{$b$}
\FAProp(11.,12.95)(5.5,10.)(0.,){/Straight}{1}
\FALabel(7.99737,12.3609)[br]{$t$}
\FAProp(11.,12.95)(15.,15.)(0.,){/ScalarDash}{1}
\FALabel(12.7731,14.8744)[br]{$\phi$}
\FAProp(15.,15.)(15.,5.)(0.,){/Sine}{1}
\FALabel(16.07,10.)[l]{$W$}
\FAVert(11.,12.95){0}
\FAVert(11.,7.1){0}
\FAVert(5.5,10.){0}
\FAVert(15.,15.){0}
\FAVert(15.,5.){0}

\FADiagram{}
\FAProp(0.,10.)(5.5,10.)(0.,){/ScalarDash}{0}
\FALabel(2.75,9.18)[t]{$H$}
\FAProp(20.,15.)(15.,15.)(0.,){/Sine}{0}
\FALabel(17.5,16.07)[b]{$\gamma$}
\FAProp(20.,5.)(15.,5.)(0.,){/Sine}{0}
\FALabel(17.5,3.93)[t]{$\gamma$}
\FAProp(11.,7.1)(5.5,10.)(0.,){/ScalarDash}{-1}
\FALabel(8.00707,7.65892)[tr]{$\phi$}
\FAProp(11.,7.1)(15.,5.)(0.,){/Straight}{-1}
\FALabel(12.7595,5.15763)[tr]{$t$}
\FAProp(11.,12.95)(11.,7.1)(0.,){/Straight}{-1}
\FALabel(12.07,10.025)[l]{$b$}
\FAProp(11.,12.95)(5.5,10.)(0.,){/ScalarDash}{1}
\FALabel(7.99737,12.3609)[br]{$\phi$}
\FAProp(11.,12.95)(15.,15.)(0.,){/Straight}{1}
\FALabel(12.7731,14.8744)[br]{$t$}
\FAProp(15.,15.)(15.,5.)(0.,){/Straight}{1}
\FALabel(16.07,10.)[l]{$t$}
\FAVert(11.,12.95){0}
\FAVert(11.,7.1){0}
\FAVert(5.5,10.){0}
\FAVert(15.,15.){0}
\FAVert(15.,5.){0}

\FADiagram{}
\FAProp(0.,10.)(5.5,10.)(0.,){/ScalarDash}{0}
\FALabel(2.75,9.18)[t]{$H$}
\FAProp(20.,15.)(15.,15.)(0.,){/Sine}{0}
\FALabel(17.5,16.07)[b]{$\gamma$}
\FAProp(20.,5.)(15.,5.)(0.,){/Sine}{0}
\FALabel(17.5,3.93)[t]{$\gamma$}
\FAProp(11.,7.1)(5.5,10.)(0.,){/Sine}{-1}
\FALabel(8.00707,7.65892)[tr]{$W$}
\FAProp(11.,7.1)(15.,5.)(0.,){/Straight}{-1}
\FALabel(12.7595,5.15763)[tr]{$t$}
\FAProp(11.,12.95)(11.,7.1)(0.,){/Straight}{-1}
\FALabel(12.07,10.025)[l]{$b$}
\FAProp(11.,12.95)(5.5,10.)(0.,){/ScalarDash}{1}
\FALabel(7.99737,12.3609)[br]{$\phi$}
\FAProp(11.,12.95)(15.,15.)(0.,){/Straight}{1}
\FALabel(12.7731,14.8744)[br]{$t$}
\FAProp(15.,15.)(15.,5.)(0.,){/Straight}{1}
\FALabel(16.07,10.)[l]{$t$}
\FAVert(11.,12.95){0}
\FAVert(11.,7.1){0}
\FAVert(5.5,10.){0}
\FAVert(15.,15.){0}
\FAVert(15.,5.){0}

\FADiagram{}
\FAProp(0.,10.)(5.5,10.)(0.,){/ScalarDash}{0}
\FALabel(2.75,9.18)[t]{$H$}
\FAProp(20.,15.)(15.,15.)(0.,){/Sine}{0}
\FALabel(17.5,16.07)[b]{$\gamma$}
\FAProp(20.,5.)(15.,5.)(0.,){/Sine}{0}
\FALabel(17.5,3.93)[t]{$\gamma$}
\FAProp(5.5,10.)(15.,15.)(0.,){/Straight}{1}
\FALabel(10.0081,13.3916)[br]{$t$}
\FAProp(9.7,7.75)(5.5,10.)(0.,){/Straight}{1}
\FALabel(7.34806,7.98871)[tr]{$t$}
\FAProp(9.7,7.75)(15.,5.)(0.,){/ScalarDash}{1}
\FALabel(12.1161,5.47918)[tr]{$\phi$}
\FAProp(15.,10.5)(9.7,7.75)(0.,){/Straight}{1}
\FALabel(12.1161,10.0208)[br]{$b$}
\FAProp(15.,10.5)(15.,15.)(0.,){/Straight}{-1}
\FALabel(16.07,12.75)[l]{$t$}
\FAProp(15.,10.5)(15.,5.)(0.,){/ScalarDash}{-1}
\FALabel(16.07,7.75)[l]{$\phi$}
\FAVert(15.,10.5){0}
\FAVert(9.7,7.75){0}
\FAVert(5.5,10.){0}
\FAVert(15.,15.){0}
\FAVert(15.,5.){0}

\FADiagram{}
\FAProp(0.,10.)(5.5,10.)(0.,){/ScalarDash}{0}
\FALabel(2.75,9.18)[t]{$H$}
\FAProp(20.,15.)(15.,15.)(0.,){/Sine}{0}
\FALabel(17.5,16.07)[b]{$\gamma$}
\FAProp(20.,5.)(15.,5.)(0.,){/Sine}{0}
\FALabel(17.5,3.93)[t]{$\gamma$}
\FAProp(5.5,10.)(15.,15.)(0.,){/ScalarDash}{1}
\FALabel(10.0081,13.3916)[br]{$\phi$}
\FAProp(9.7,7.75)(5.5,10.)(0.,){/ScalarDash}{1}
\FALabel(7.34806,7.98871)[tr]{$\phi$}
\FAProp(9.7,7.75)(15.,5.)(0.,){/Straight}{-1}
\FALabel(12.1161,5.47918)[tr]{$b$}
\FAProp(15.,10.5)(9.7,7.75)(0.,){/Straight}{-1}
\FALabel(12.1161,10.0208)[br]{$t$}
\FAProp(15.,10.5)(15.,15.)(0.,){/Sine}{-1}
\FALabel(16.07,12.75)[l]{$W$}
\FAProp(15.,10.5)(15.,5.)(0.,){/Straight}{1}
\FALabel(16.07,7.75)[l]{$b$}
\FAVert(15.,10.5){0}
\FAVert(9.7,7.75){0}
\FAVert(5.5,10.){0}
\FAVert(15.,15.){0}
\FAVert(15.,5.){0}

\FADiagram{}
\FAProp(0.,10.)(4.5,10.)(0.,){/ScalarDash}{0}
\FALabel(2.25,9.18)[t]{$H$}
\FAProp(20.,15.)(14.6,10.)(0.,){/Sine}{0}
\FALabel(17.8731,11.8427)[tl]{$\gamma$}
\FAProp(20.,5.)(14.6,10.)(0.,){/Sine}{0}
\FALabel(16.7269,6.84267)[tr]{$\gamma$}
\FAProp(7.,13.)(4.5,10.)(0.211476,){/ScalarDash}{-1}
\FALabel(4.72204,12.2766)[br]{$\phi$}
\FAProp(12.,13.)(7.,13.)(0.8,){/Straight}{1}
\FALabel(9.5,16.07)[b]{$t$}
\FAProp(12.,13.)(7.,13.)(-0.8,){/Straight}{-1}
\FALabel(9.5,9.93)[t]{$b$}
\FAProp(12.,13.)(14.6,10.)(-0.182242,){/ScalarDash}{1}
\FALabel(14.2647,12.2721)[bl]{$\phi$}
\FAProp(4.5,10.)(14.6,10.)(0.693069,){/ScalarDash}{-1}
\FALabel(9.55,5.43)[t]{$\phi$}
\FAVert(12.,13.){0}
\FAVert(7.,13.){0}
\FAVert(4.5,10.){0}
\FAVert(14.6,10.){0}

\FADiagram{}
\FAProp(0.,10.)(5.5,10.)(0.,){/ScalarDash}{0}
\FALabel(2.75,9.18)[t]{$H$}
\FAProp(20.,15.)(14.5,10.)(0.,){/Sine}{0}
\FALabel(17.6874,14.1669)[br]{$\gamma$}
\FAProp(20.,5.)(14.5,10.)(0.,){/Sine}{0}
\FALabel(17.8126,8.16691)[bl]{$\gamma$}
\FAProp(10.,14.5)(5.5,10.)(0.4,){/Straight}{-1}
\FALabel(6.23398,13.766)[br]{$t$}
\FAProp(10.,5.5)(10.,14.5)(0.,){/Straight}{-1}
\FALabel(11.07,10.)[l]{$b$}
\FAProp(10.,5.5)(5.5,10.)(-0.4,){/Straight}{1}
\FALabel(6.23398,6.23398)[tr]{$t$}
\FAProp(10.,5.5)(14.5,10.)(0.4,){/Sine}{1}
\FALabel(13.766,6.23398)[tl]{$W$}
\FAProp(10.,14.5)(14.5,10.)(-0.4,){/Sine}{-1}
\FALabel(12.266,14.366)[bl]{$W$}
\FAVert(10.,5.5){0}
\FAVert(10.,14.5){0}
\FAVert(5.5,10.){0}
\FAVert(14.5,10.){0}

\FADiagram{}
\FAProp(0.,10.)(8.,6.)(0.,){/ScalarDash}{0}
\FALabel(3.89862,7.31724)[tr]{$H$}
\FAProp(20.,15.)(13.,15.)(0.,){/Sine}{0}
\FALabel(16.5,16.07)[b]{$\gamma$}
\FAProp(20.,5.)(8.,6.)(0.,){/Sine}{0}
\FALabel(13.8713,4.43535)[t]{$\gamma$}
\FAProp(9.2,13.9)(13.,15.)(-0.150337,){/ScalarDash}{-1}
\FALabel(10.5863,15.7445)[b]{$\phi$}
\FAProp(7.,9.5)(9.2,13.9)(0.8,){/Straight}{1}
\FALabel(10.7664,10.6068)[tl]{$t$}
\FAProp(7.,9.5)(9.2,13.9)(-0.8,){/Straight}{-1}
\FALabel(5.43364,12.7932)[br]{$b$}
\FAProp(7.,9.5)(8.,6.)(0.222749,){/ScalarDash}{1}
\FALabel(6.13962,9.00132)[r]{$\phi$}
\FAProp(13.,15.)(8.,6.)(-0.605305,){/Sine}{-1}
\FALabel(14.0988,8.71399)[tl]{$W$}
\FAVert(7.,9.5){0}
\FAVert(9.2,13.9){0}
\FAVert(13.,15.){0}
\FAVert(8.,6.){0}

\FADiagram{}
\FAProp(0.,10.)(5.5,10.)(0.,){/ScalarDash}{0}
\FALabel(2.75,9.18)[t]{$H$}
\FAProp(20.,15.)(10.,10.)(0.,){/Sine}{0}
\FALabel(17.1318,12.7564)[tl]{$\gamma$}
\FAProp(20.,5.)(14.5,10.)(0.,){/Sine}{0}
\FALabel(17.8126,8.16691)[bl]{$\gamma$}
\FAProp(10.,5.5)(5.5,10.)(-0.4,){/Straight}{1}
\FALabel(6.23398,6.23398)[tr]{$t$}
\FAProp(10.,5.5)(10.,10.)(0.,){/Straight}{-1}
\FALabel(11.07,7.75)[l]{$b$}
\FAProp(10.,5.5)(14.5,10.)(0.4,){/ScalarDash}{1}
\FALabel(13.766,6.23398)[tl]{$\phi$}
\FAProp(10.,14.5)(5.5,10.)(0.4,){/Straight}{-1}
\FALabel(6.23398,13.766)[br]{$t$}
\FAProp(10.,14.5)(10.,10.)(0.,){/Straight}{1}
\FALabel(8.93,12.25)[r]{$b$}
\FAProp(10.,14.5)(14.5,10.)(-0.4,){/Sine}{-1}
\FALabel(13.1968,14.1968)[bl]{$W$}
\FAVert(10.,14.5){0}
\FAVert(10.,5.5){0}
\FAVert(5.5,10.){0}
\FAVert(10.,10.){0}
\FAVert(14.5,10.){0}

\FADiagram{}
\FAProp(0.,10.)(5.5,10.)(0.,){/ScalarDash}{0}
\FALabel(2.75,9.18)[t]{$H$}
\FAProp(20.,15.)(14.5,15.)(0.,){/Sine}{0}
\FALabel(17.25,16.07)[b]{$\gamma$}
\FAProp(20.,5.)(14.5,5.)(0.,){/Sine}{0}
\FALabel(17.25,3.93)[t]{$\gamma$}
\FAProp(5.5,10.)(14.5,15.)(0.,){/Sine}{-1}
\FALabel(9.72725,13.3749)[br]{$W$}
\FAProp(5.5,10.)(14.5,5.)(0.,){/Sine}{1}
\FALabel(9.72725,6.62506)[tr]{$W$}
\FAProp(14.5,12.5)(14.5,15.)(0.,){/ScalarDash}{1}
\FALabel(15.57,13.30)[l]{$\phi$}
\FAProp(14.5,7.5)(14.5,12.5)(0.8,){/Straight}{-1}
\FALabel(17.57,10.)[l]{$t$}
\FAProp(14.5,7.5)(14.5,12.5)(-0.8,){/Straight}{1}
\FALabel(11.43,10.)[r]{$b$}
\FAProp(14.5,7.5)(14.5,5.)(0.,){/Sine}{-1}
\FALabel(15.57,6.50)[l]{$W$}
\FAVert(14.5,7.5){0}
\FAVert(14.5,12.5){0}
\FAVert(5.5,10.){0}
\FAVert(14.5,15.){0}
\FAVert(14.5,5.){0}

\FADiagram{}
\FAProp(0.,10.)(6.5,10.)(0.,){/ScalarDash}{0}
\FALabel(3.25,9.18)[t]{$H$}
\FAProp(20.,15.)(13.,14.)(0.,){/Sine}{0}
\FALabel(16.2808,15.5544)[b]{$\gamma$}
\FAProp(20.,5.)(13.,6.)(0.,){/Sine}{0}
\FALabel(16.2808,4.44558)[t]{$\gamma$}
\FAProp(6.5,10.)(13.,14.)(0.,){/ScalarDash}{1}
\FALabel(9.20801,13.1807)[br]{$\phi$}
\FAProp(6.5,10.)(13.,6.)(0.,){/ScalarDash}{-1}
\FALabel(9.20801,6.81927)[tr]{$\phi$}
\FAProp(13.,14.)(13.,10.)(0.,){/ScalarDash}{1}
\FALabel(10.274,10.)[l]{$\phi$}
\FAProp(13.,10.)(13.,6.)(0.,){/ScalarDash}{1}
\FAProp(13.,10.)(16.,10.)(0.,){/ScalarDash}{0}
\FALabel(13.500,11.7)[l]{$H$}
\FAProp(16.,10.)(16.,10.)(20.,10.){/Straight}{1}
\FALabel(17.674,10.)[l]{$t$}
\FAVert(6.5,10.){0}
\FAVert(13.,14.){0}
\FAVert(13.,6.){0}
\FAVert(16.,10.){0}
\FAVert(13.,10.){0}

\FADiagram{}
\FAProp(0.,10.)(6.5,10.)(0.,){/ScalarDash}{0}
\FALabel(3.25,9.18)[t]{$H$}
\FAProp(20.,15.)(13.,14.)(0.,){/Sine}{0}
\FALabel(16.2808,15.5544)[b]{$\gamma$}
\FAProp(20.,5.)(13.,6.)(0.,){/Sine}{0}
\FALabel(16.2808,4.44558)[t]{$\gamma$}
\FAProp(6.5,10.)(13.,14.)(0.,){/GhostDash}{1}
\FALabel(9.20801,13.1807)[br]{$u_-$}
\FAProp(6.5,10.)(13.,6.)(0.,){/GhostDash}{-1}
\FALabel(9.20801,6.81927)[tr]{$u_-$}
\FAProp(13.,14.)(13.,10.)(0.,){/GhostDash}{1}
\FALabel(9.474,9.7)[l]{$u_-$}
\FAProp(13.,10.)(13.,6.)(0.,){/GhostDash}{1}
\FAProp(13.,10.)(16.,10.)(0.,){/ScalarDash}{0}
\FALabel(13.500,11.7)[l]{$H$}
\FAProp(16.,10.)(16.,10.)(20.,10.){/Straight}{1}
\FALabel(17.674,10.)[l]{$t$}
\FAVert(6.5,10.){0}
\FAVert(13.,14.){0}
\FAVert(13.,6.){0}
\FAVert(16.,10.){0}
\FAVert(13.,10.){0}

\FADiagram{}
\FAProp(0.,10.)(6.5,10.)(0.,){/ScalarDash}{0}
\FALabel(3.25,12.5)[t]{$H$}
\FAProp(6.5,10.)(4.5,7.)(0.,){/ScalarDash}{0}
\FALabel(4.80,8.5)[r]{$H$}
\FAProp(4.5,7.)(4.5,7.)(2.5,4.){/Straight}{1}
\FALabel(2.8,5.3)[l]{$t$}
\FAProp(20.,15.)(13.,14.)(0.,){/Sine}{0}
\FALabel(16.2808,15.5544)[b]{$\gamma$}
\FAProp(20.,5.)(13.,6.)(0.,){/Sine}{0}
\FALabel(16.2808,4.44558)[t]{$\gamma$}
\FAProp(6.5,10.)(13.,14.)(0.,){/Sine}{1}
\FALabel(9.20801,13.1807)[br]{$W$}
\FAProp(6.5,10.)(13.,6.)(0.,){/Sine}{-1}
\FALabel(9.90801,6.81927)[tr]{$W$}
\FAProp(13.,14.)(13.,6.)(0.,){/Sine}{1}
\FALabel(14.274,10.)[l]{$W$}
\FAVert(6.5,10.){0}
\FAVert(13.,14.){0}
\FAVert(13.,6.){0}
\FAVert(6.5,10.){0}
\FAVert(4.5,7.){0}

\FADiagram{}
\FAProp(0.,10.)(6.5,10.)(0.,){/ScalarDash}{0}
\FALabel(3.25,9.18)[t]{$H$}
\FAProp(20.,15.)(13.,14.)(0.,){/Sine}{0}
\FALabel(16.2808,15.5544)[b]{$\gamma$}
\FAProp(13.,14.)(10.,16.)(0.,){/ScalarDash}{0}
\FALabel(11.20,14.)[r]{$H$}
\FAProp(10.,16.)(10.,16.)(7.,18.){/Straight}{1}
\FALabel(8.0,17.)[l]{$t$}
\FAProp(20.,5.)(13.,6.)(0.,){/Sine}{0}
\FALabel(16.2808,4.44558)[t]{$\gamma$}
\FAProp(6.5,10.)(13.,14.)(0.,){/ScalarDash}{-1}
\FALabel(7.00801,10.9807)[br]{$\phi$}
\FAProp(6.5,10.)(13.,6.)(0.,){/ScalarDash}{1}
\FALabel(9.20801,6.81927)[tr]{$\phi$}
\FAProp(13.,14.)(13.,6.)(0.,){/Sine}{-1}
\FALabel(14.274,10.)[l]{$W$}
\FAVert(6.5,10.){0}
\FAVert(13.,14.){0}
\FAVert(13.,6.){0}
\FAVert(13.,14.){0}
\FAVert(10.,16.){0}

\end{feynartspicture}

\caption{\label{2loop} Sample Feynman diagrams contributing at the two-loop
electroweak order to the process $H\to\gamma\gamma$.}

\end{center}
\end{figure}

The contributions of $\mathcal{O}(G_F M_t^2)$ are obtained by considering all
two-loop electroweak diagrams involving a virtual top quark.
This includes also the tadpole diagrams with a closed top-quark loop, which
are proportional to $M_t^4$.
For arbitrary gauge parameter, this leads us to consider a total of
$\mathcal{O}(1000)$ diagrams.
Some of them are depicted in Fig.~\ref{2loop}.
These diagrams naturally split into two classes.
The first class consists of those diagrams where a neutral boson, i.e.\ a
Higgs boson or a neutral Goldstone boson ($\chi$), is added to the one-loop
top-quark diagrams.
The exchange of a $Z$ boson does not produce quadratic contributions in $M_t$.
The application of the asymptotic-expansion technique to these diagrams leads
to a simple Taylor expansion in the external momenta.

\begin{figure}[th]
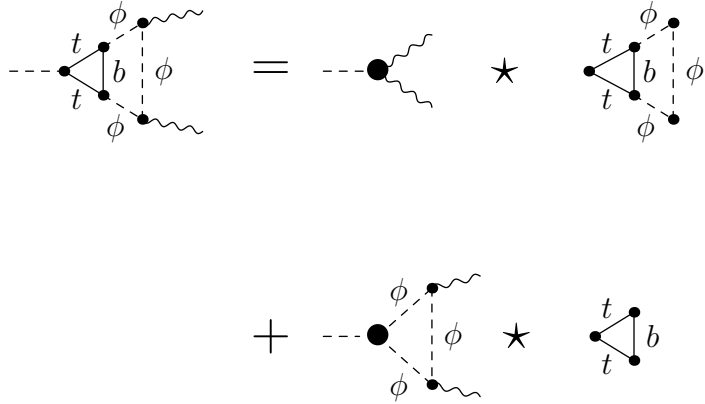

\begin{center}

\unitlength=1bp%

\begin{feynartspicture}(370,200)(3,2)

\FADiagram{}
\FAProp(0.,10.)(4.5,10.)(0.,){/ScalarDash}{0}
\FAProp(16.,15.)(11.,14.)(0.,){/Sine}{0}
\FAProp(16.,5.)(11.,6.)(0.,){/Sine}{0}
\FAProp(4.5,10.)(7.75,12.)(0.,){/Straight}{0}
\FALabel(6.,11.5)[br]{$t$}
\FAProp(4.5,10.)(7.75,8.)(0.,){/Straight}{0}
\FALabel(6.,8.5)[tr]{$t$}
\FAProp(7.75,8.)(11.,6.)(0.,){/ScalarDash}{0}
\FALabel(9.5,6.5)[tr]{$\phi$}
\FAProp(11.,6.)(11.,14.)(0.,){/ScalarDash}{0}
\FALabel(12.,10.)[l]{$\phi$}
\FAProp(7.75,12.)(11.,14.)(0.,){/ScalarDash}{0}
\FALabel(9.7,13.5)[br]{$\phi$}
\FAProp(7.75,12.)(7.75,8.)(0.,){/Straight}{0}
\FALabel(8.5,10.)[l]{$b$}
\FAVert(4.5,10.){0}
\FAVert(11.,14.){0}
\FAVert(11.,6.){0}
\FAVert(7.75,8.){0}
\FAVert(7.75,12.){0}
\FALabel(20.,10.)[l]{\LARGE{=}}

\FADiagram{}
\FAProp(4.,10.)(8.5,10.)(0.,){/ScalarDash}{0}
\FAProp(8.5,10.)(13.,13.)(0.,){/Sine}{0}
\FAProp(8.5,10.)(13.,7.)(0.,){/Sine}{0}
\FAVert(8.5,10.){0}
\FALabel(8.5,10.)[]{\LARGE{$\bullet$}}
\FALabel(19,10.)[]{\LARGE{$\star$}}

\FADiagram{}
\FAProp(4.,10.)(7.75,12.)(0.,){/Straight}{0}
\FALabel(6.,11.5)[br]{$t$}
\FAProp(4.,10.)(7.75,8.)(0.,){/Straight}{0}
\FALabel(6.,8.5)[tr]{$t$}
\FAProp(7.75,8)(11.,6.)(0.,){/ScalarDash}{0}
\FALabel(9.5,6.5)[tr]{$\phi$}
\FAProp(11.,6.)(11.,14.)(0.,){/ScalarDash}{0}
\FALabel(12.,10.)[l]{$\phi$}
\FAProp(7.75,12.)(11.,14.)(0.,){/ScalarDash}{0}
\FALabel(9.7,13.5)[br]{$\phi$}
\FAProp(7.75,12.)(7.75,8.)(0.,){/Straight}{0}
\FALabel(8.5,10.)[l]{$b$}
\FAVert(7.75,8.){0}
\FAVert(7.75,12.){0}
\FAVert(4.,10.){0}
\FAVert(11.,14.){0}
\FAVert(11.,6.){0}

\FADiagram{}
\FALabel(20.,10.)[l]{\LARGE{+}}

\FADiagram{}
\FAProp(4.,10.)(7.,10.)(0.,){/ScalarDash}{0}
\FAProp(13.,14.)(17.,15.)(0.,){/Sine}{0}
\FAProp(13.,6.)(17.,5.)(0.,){/Sine}{0}
\FAProp(8.5,10.)(13.,14.)(0.,){/ScalarDash}{0}
\FALabel(11.,12.5)[rb]{$\phi$}
\FAProp(8.5,10.)(13.,6.)(0.,){/ScalarDash}{0}
\FALabel(11.,7.)[rt]{$\phi$}
\FAProp(13.,14.)(13.,6.)(0.,){/ScalarDash}{0}
\FALabel(14.,10.)[l]{$\phi$}
\FAVert(8.5,10.){0}
\FAVert(13.,14.){0}
\FAVert(13.,6.){0}
\FALabel(8.5,10.)[]{\LARGE{$\bullet$}}
\FALabel(20.,10.)[]{\LARGE{$\star$}}

\FADiagram{}
\FAProp(4.5,10.)(7.75,12.)(0.,){/Straight}{0}
\FALabel(6.,11.5)[br]{$t$}
\FAProp(4.5,10.)(7.75,8.)(0.,){/Straight}{0}
\FALabel(6.,8.5)[tr]{$t$}
\FAProp(7.75,8.)(7.75,12.)(0.,){/Straight}{0}
\FALabel(8.75,10.)[l]{$b$}
\FAVert(4.5,10.){0}
\FAVert(7.75,12.){0}
\FAVert(7.75,8.){0}

\end{feynartspicture}

\caption{\label{Mt4AsEx} Diagrammatic asymptotic expansion of a Feynman
diagram that produces $M_{t}^{4}$ terms.}

\end{center}
\end{figure}

This is different for the second class of diagrams, which, next to the top
quark, also contain a $W$ or $\phi$ boson and, as a consequence, also the
bottom quark, which we take to be massless throughout the calculation. 
Due to the presence of cuts through light-particle lines, the
asymptotic-expansion technique applied to these diagrams also yields
nontrivial terms, as is exemplified in Fig.~\ref{Mt4AsEx}.
The first contribution on the right-hand side of the equation in 
Fig.~\ref{Mt4AsEx} symbolizes the naive Taylor expansion in the external
momenta.
In the second contribution, the subdiagram to the right of the star has to be
expanded in its external momenta, which also includes the loop momentum of the
co-subgraph to the left of the star.
The result of the expansion is inserted as an effective vertex, and the
remaining integration over the second loop-momentum is performed after a
further expansion in $q_1$ and $q_2$.
The latter is allowed, since we work in the limit $(2M_W)^2 \gg M_H^2$.
We wish to mention that such contributions develop $M_t^4$ terms, which cancel
in the final result only in combination with the $M_t^4$ terms from the
counterterms of Eq.~(\ref{eq:ct}) and the genuine two-loop tadpole diagrams.
For this reason, it is crucial to include the latter in our calculation.
As a further comment, we note that, unlike the example of Fig.~\ref{Mt4AsEx},
it can also happen that the co-subgraph only involves massless bottom quarks,
and no expansion in the external momenta is allowed.
In our calculation, the contributions from such co-subgraphs vanish.


As already mentioned above, we use in our calculation a general gauge
parameter $\xi_W$ related to the $W$ boson and verify that our final result is
independent of $\xi_W$.
We do this in the limit of large and small values of $\xi_W$.
To this end, we apply the asymptotic-expansion technique in the following four
limiting cases
\begin{eqnarray}
  (i)   && M_{t}^2 \gg M_{W}^2 =   \xi_W M_{W}^2 \gg M_H^2,
  \nonumber\\
  (ii)  && M_{t}^2 \gg M_{W}^2 \gg \xi_W M_{W}^2 \gg M_H^2,
  \nonumber\\
  (iii) && M_{t}^2 \gg \xi_W M_{W}^2 \gg M_{W}^2 \gg M_H^2,
  \nonumber\\
  (iv)  && \xi_W M_{W}^2 \gg M_{t}^2 \gg M_{W}^2 \gg M_H^2,
\end{eqnarray}
where the inequality $M_{W}^2 \gg M_H^2$ has to be understood in a formal
sense, as, in practice, one has $(2M_W)^2\gg M_H^2$.
The result we obtain by asymptotic expansion is an expansion of the exact
result for the $\mathcal{O}(G_{F} M_{t}^{2})$ contribution in powers of
$\tau_{W}$.
In case $(i)$, where $\xi_W=1$, which corresponds to 't~Hooft-Feynman gauge,
we are able to evaluate the first five terms of this expansion.
In cases $(ii)$--$(iv)$, we compute the first two expansion terms and find
agreement with the result obtained for $\xi_W=1$.

Our final result for ${\cal A}_{tW}^{(1)}$ emerges as the sum
\begin{equation}
  \mathcal{A}_{tW}^{(1)} = 
  \mathcal{A}_{u}^{(1)} +
  \mathcal{A}_{H,\chi}^{(1)} +
  \mathcal{A}_{W,\phi}^{(1)},
\label{eq:sum}
\end{equation}
where $\mathcal{A}_{u}^{(1)}$ is the universal contribution induced by the
one-loop term $\delta_u$ of Eq.~(\ref{RenUniv}),
$\mathcal{A}_{H,\chi}^{(1)}$ is the two-loop contribution involving 
virtual $H$ and $\chi$ bosons, and $\mathcal{A}_{W,\phi}^{(1)}$ the
remaining two-loop contribution involving virtual $W$ and $\phi$ bosons.
In $\mathcal{A}_{H,\chi}^{(1)}$ and $\mathcal{A}_{W,\phi}^{(1)}$, also
the corresponding counterterm and tadpole contributions are included.
For the individual pieces, we find
\begin{eqnarray}
  \mathcal{A}_{u}^{(1)} & = & 
  \hat{\mathcal{A}} N_c x_t \left( - \frac{329}{108} 
    - \frac{77}{90} \tau_{W} - \frac{19}{45} \tau_{W}^{2} 
    - \frac{58}{225} \tau_{W}^{3} - \frac{1312}{7425} \tau_{W}^{4} 
    +\cdots
  \right) ,
  \nonumber\\
  \mathcal{A}_{H,\chi}^{(1)} & = & 
  \hat{\mathcal{A}} N_c x_t \left( - \frac{8}{27} \right) ,
  \nonumber\\
  \mathcal{A}_{W,\phi}^{(1)} & = & 
  \hat{\mathcal{A}} N_c x_t \left( \frac{182}{27} + \frac{22}{15} \tau_{W} +
    \frac{76}{105} \tau_{W}^{2} + \frac{232}{525} \tau_{W}^{3} 
    + \frac{5248}{17325} \tau_{W}^{4} +\cdots
  \right) ,
\label{eq:terms}
\end{eqnarray}
where $\hat{\mathcal{A}}$ is defined below Eq.~(\ref{eq::born}) and the
ellipses indicate terms of $\mathcal{O}(\tau_W^5)$.
Notice that the leading $\mathcal{O}(G_F M_t^2)$ term of
$\mathcal{A}_{H,\chi}^{(1)}$ is not accompanied by an expansion in $\tau_W$,
since the contributing diagrams do not involve virtual $W$ or $\phi$ bosons.
On the other hand, detailed inspection reveals that there is also no 
expansion in the parameter $M_H^2/(2M_Z)^2$, contrary to what might be
expected at first sight.
Inserting Eq.~(\ref{eq:terms}) into Eq.~(\ref{eq:sum}), we obtain our final
result
\begin{equation}
  \mathcal{A}_{tW}^{(1)} = 
  \hat{\mathcal{A}} N_c x_t \left( \frac{367}{108} + \frac{11}{18} \tau_{W} +
  \frac{19}{63} \tau_{W}^{2} + \frac{58}{315} \tau_{W}^{3} 
  + \frac{1312}{10395} \tau_{W}^{4} +\cdots
\right).
\end{equation}

The correction of $\mathcal{O}(G_F M_t^2)$ to $\Gamma(H\to\gamma\gamma)$ was
also considered in Ref.~\cite{Liao:1996td}.
The expression found in that reference disagrees with our result.
One reason is probably that the authors of Ref.~\cite{Liao:1996td} only
considered virtual $\phi$ bosons, but disregarded virtual $W$-bosons.
However, our calculation shows that both types of charged bosons have to be
taken into account at the same time in order to arrive at an
ultraviolet-finite and gauge-parameter-independent result.
In Ref.~\cite{Djouadi:1997rj}, the dominant two-loop electroweak corrections
to the Higgs-boson couplings to pairs of gauge bosons and light fermions
induced by a sequential isodoublet of ultraheavy quarks $(A,B)$ were
investigated by means of a low-energy theorem \cite{Shi79,let}.
In that paper, also a result for the $\mathcal{O}(G_F M_t^2)$ correction to
$\Gamma(H\to\gamma\gamma)$ is presented, which is obtained by sending the mass
of the fourth-generation down quark ($M_B$) to zero at the end of the
calculation, which is performed assuming that $M_B\gg M_W$.
This result also deviates from the one obtained above.
Detailed inspection reveals that this difference may be attributed to the
interchange of limits performed in Ref.~\cite{Djouadi:1997rj}, which is not
justified in the case under consideration, although such a procedure is known
to lead to correct results in simpler examples.
In fact, reanalyzing the limiting case $M_A\gg M_B\gg M_W$ using the
asymptotic-expansion technique, we are able to reproduce the terms of
Eqs.~(48)--(54) in Ref.~\cite{Djouadi:1997rj} that survive in this limit and,
at the same time, to identify contributions that do not occur if $M_B=0$ is
imposed at the outset of the calculation.

For completeness, we also specify our corresponding result for
$\Gamma(H\to gg)$.
Its evaluation is significantly simpler, since, at one loop, only the
top-quark diagrams contribute.
Consequently, to obtain the leading two-loop correction proportional to
$G_F M_t^2$, we only have to consider the exchange of $H$, $\chi$, and $\phi$
bosons.
This only requires a naive Taylor expansion in $q_1$ and $q_2$.
We obtain
\begin{equation}
  \mathcal{A}_{gg}^{(1)} = \hat{\mathcal{A}}_{gg} x_t \left( \frac{1}{3}
  \right),
\end{equation}
with $\hat{\mathcal{A}}_{gg} = 2^{3/4}G_{F}^{1/2}(\alpha_s/\pi)$.
This result is in agreement with Ref.~\cite{Djouadi:1994ge}, where a
low-energy theorem \cite{Shi79,let} was used.


\section{\label{sec::numerics}Numerical results}

We are now in a position to present our numerical results and to assess the
convergence properties of our expansions in $\tau_t$ and $\tau_W$.
We use the following numerical values for our input
parameters~\cite{Hagiwara:fs}: 
$G_{F}=1.16639\times10^{-5}$~GeV$^{-2}$,
$M_{W}=80.423$~GeV, and
$M_{t}=174.3$~GeV.

\begin{figure}[th]
\begin{center}
\epsfig{file=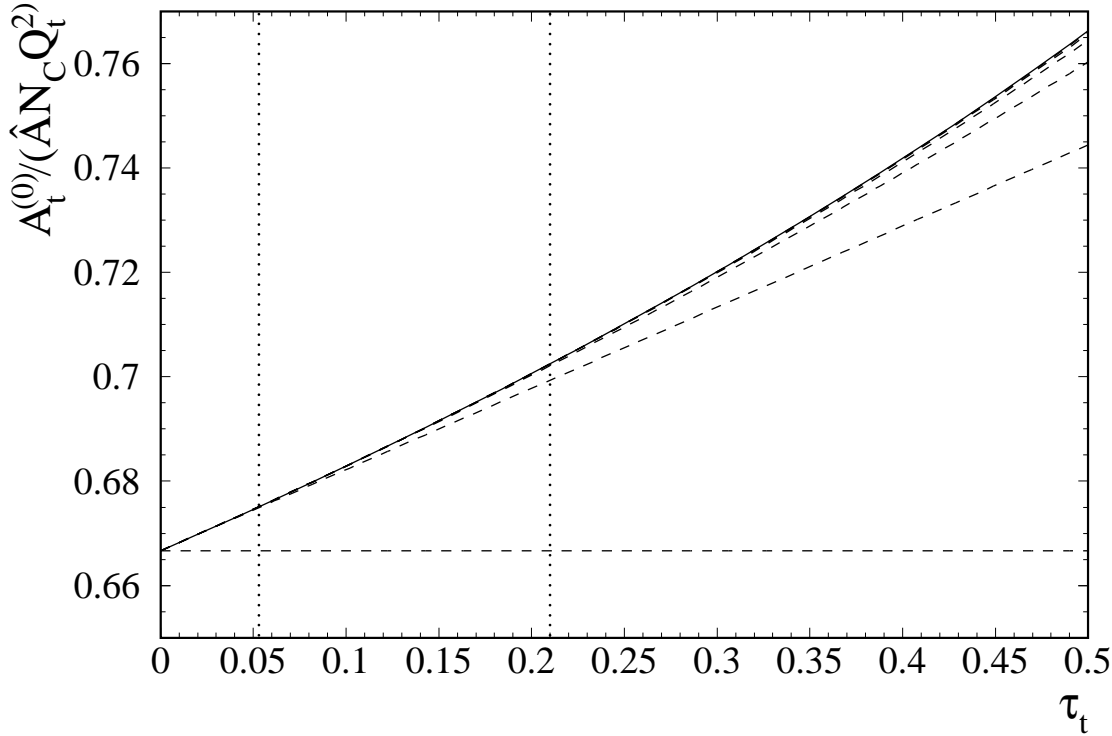,width=\textwidth}
\end{center}
\caption{\label{1loopT} 
Top-quark-induced one-loop amplitude $\mathcal{A}_t^{(0)}$ normalized to
$\hat{\mathcal{A}} N_c Q_t^2$ as a function of $\tau_t$.
The solid curve indicates the exact result, while the dashed curves represent
the sequence of approximations that are obtained by successively including
higher powers of $\tau_{t}$ in the expansion.
The dotted vertical lines encompass the intermediate-mass range of the Higgs
boson.}
\end{figure}

\begin{figure}[th]
\begin{center}
\epsfig{file=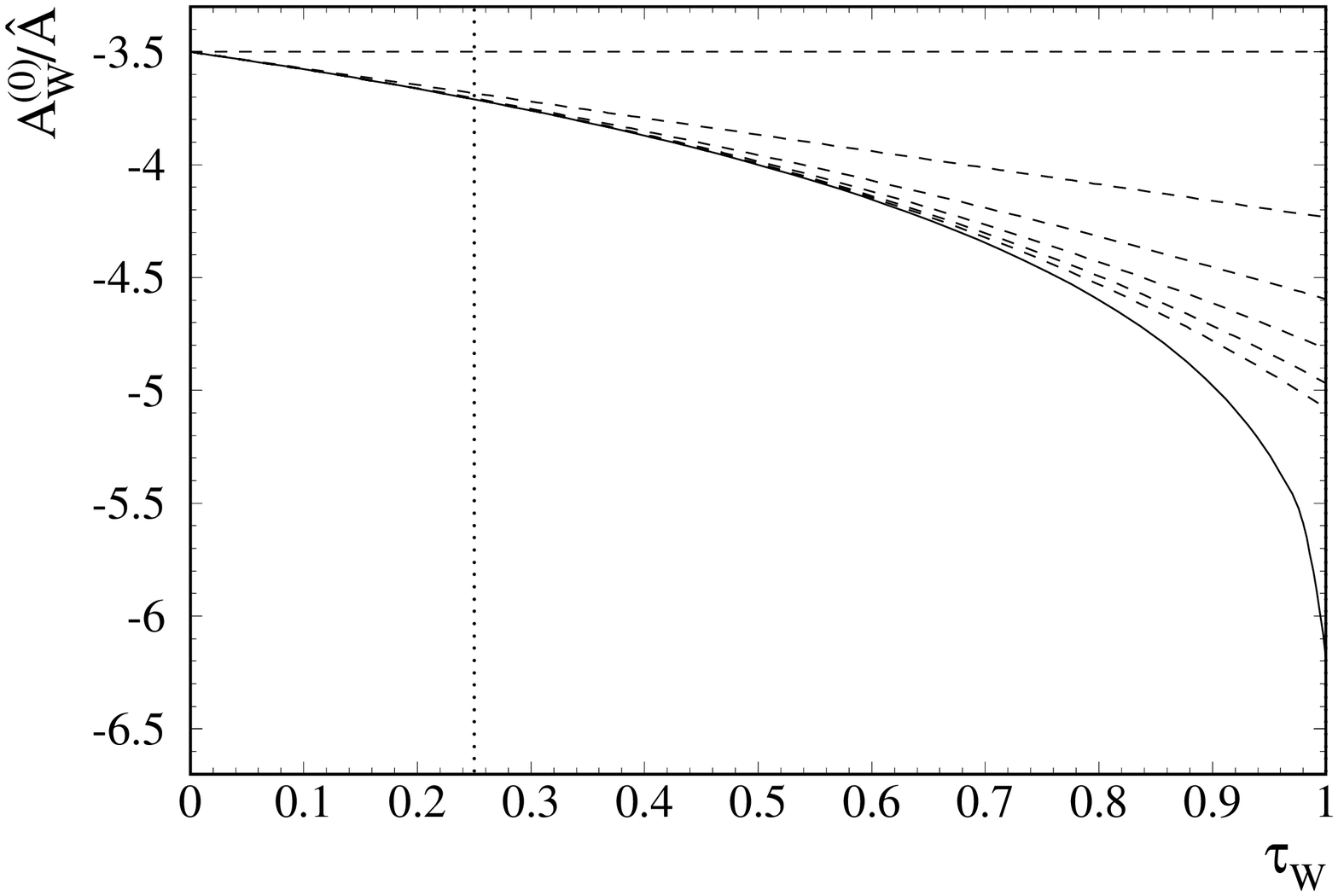,width=\textwidth}
\end{center}
\caption{\label{1loopW} 
$W$-boson-induced one-loop amplitude $\mathcal{A}_W^{(0)}$ normalized to
$\hat{\mathcal{A}}$ as a function of $\tau_W$.
The solid curve indicates the exact result, while the dashed curves represent
the sequence of approximations that are obtained by successively including
higher powers of $\tau_{W}$ in the expansion.
The dotted vertical line and the right edge of the frame encompass the
intermediate-mass range of the Higgs boson.}
\end{figure}

We first consider the one-loop amplitudes ${\cal A}_t^{(0)}$ and
${\cal A}_W^{(0)}$ induced by virtual top quarks and $W$ bosons, respectively,
for which exact results are available.
They are shown in Figs.~\ref{1loopT} and~\ref{1loopW} as functions of
$\tau_{t}$ and $\tau_{W}$, respectively.
The solid curves indicate the exact results, while the dashed curves represent
the sequences of approximations that are obtained by successively including
higher powers of $\tau_{t}$ and $\tau_{W}$, respectively, in the expansions.
The vertical lines encompass the intermediate-mass range of the Higgs boson,
$M_W\le M_H\le 2M_W$.
In Fig.~\ref{1loopW}, the second vertical line coincides with the right edge
of the frame.
From Fig.~\ref{1loopT}, we observe that the approximation consisting of the
first three terms of the expansion in $\tau_{t}$ is practically 
indistinguishable from the exact result up to $\tau_t\approx 0.25$.
The same is true for the sum of the first five expansion terms up to
$\tau_t\approx 0.5$, which corresponds to $M_H\approx 245$~GeV.
In the case of ${\cal A}_W^{(0)}$, the convergence is slightly worse, since
$M_H=2M_W$ corresponds to $\tau_W=1$ and the exact result behaves like
$\sqrt{1-\tau_{W}}$ in this limit.
Nevertheless, for $M_H=120$~GeV, 140~GeV, and $2 M_W$, the approximation by
five expansion terms deviates from the exact result by as little as
0.3\%,  1.6\%, and 19.9\%, respectively.

\begin{figure}[th]
\begin{center}
\epsfig{file=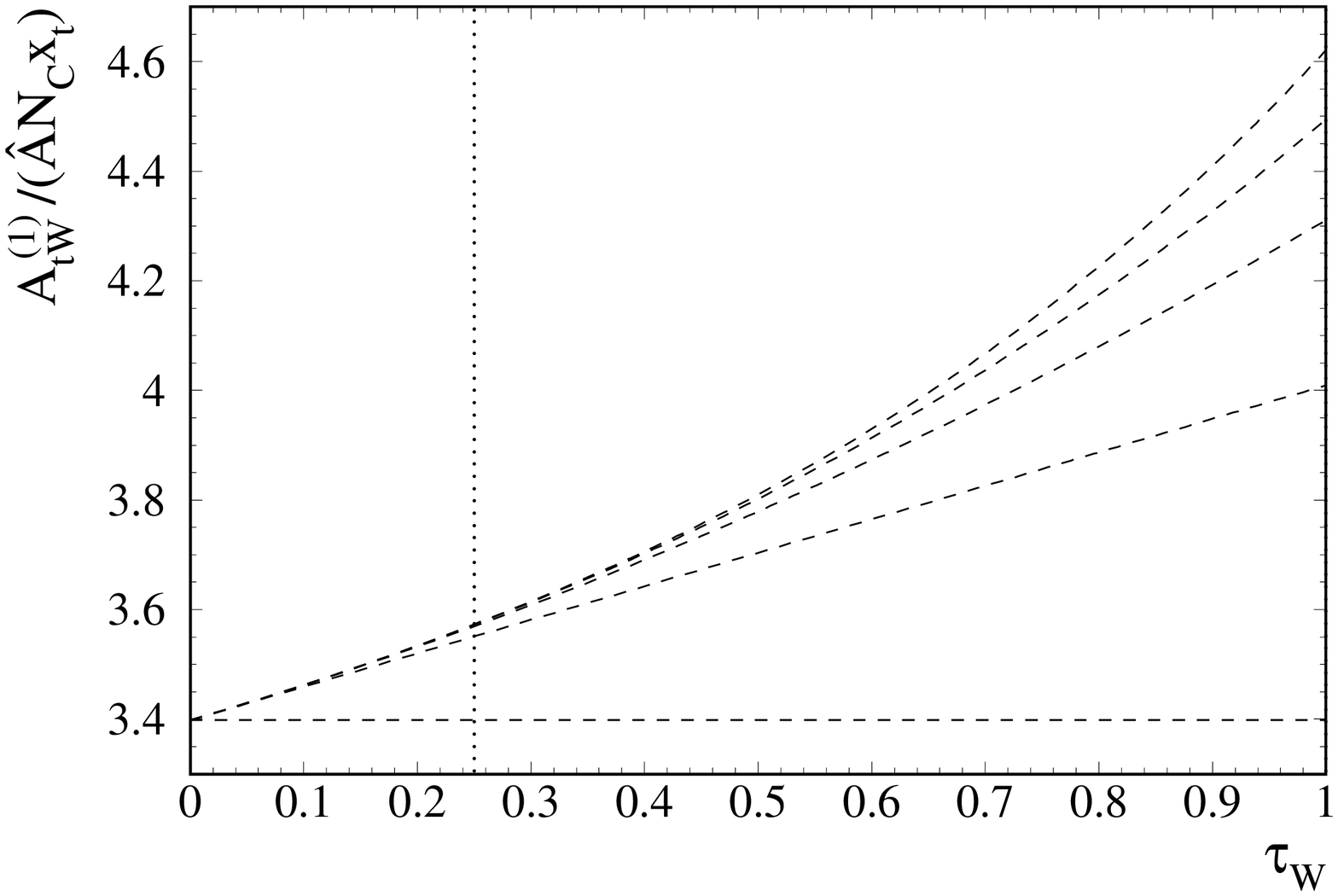,width=\textwidth}
\end{center}
\caption{\label{2loopT}
Top-quark-induced two-loop electroweak amplitude $\mathcal{A}_{tW}^{(1)}$
normalized to $\hat{\mathcal{A}} N_c x_t$ as a function of $\tau_W$.
The dashed curves represent the sequence of approximations that are obtained 
by successively including higher powers of $\tau_{W}$ in the expansion.
The dotted vertical line and the right edge of the frame encompass the
intermediate-mass range of the Higgs boson.}
\end{figure}

Having demonstrated the fast convergence of the expansions in $\tau_{t}$ and
$\tau_{W}$ of the one-loop amplitudes $\mathcal{A}_{t}^{(0)}$ and
$\mathcal{A}_{W}^{(0)}$, respectively, we now proceed to the two-loop
electroweak amplitude $\mathcal{A}_{tW}^{(1)}$, for which we found the leading
$\mathcal{O}(G_F M_t^2)$ term together with its subleading mass corrections
through $\mathcal{O}(\tau_{W}^4)$.
This corresponds to a first approximation and four improvements, which are
visualized by the five dashed curves in Fig.~\ref{2loopT}.
As in Fig.~\ref{1loopW}, the dotted vertical line and the right edge of the
frame enclose the intermediate-mass range of the Higgs boson.
We again observe rapid convergence.
The goodness of our best approximation may be assessed by considering its
relative deviation from the second best one.
For $M_H=120$~GeV, 140~GeV, and $2 M_W$, this amounts to 0.3\%, 1.0\%, and
2.8\%, respectively.
The situation is very similar to the one in Fig.~\ref{1loopW}.
In fact, the corresponding figures for $\mathcal{A}_{W}^{(0)}$ are 0.4\%,
1.1\%, and 3.1\%.
We thus expect that the goodness of the approximation of
$\mathcal{A}_{tW}^{(1)}$ by the expansion through $\mathcal{O}(\tau_{W}^4)$ is
comparable to the case of $\mathcal{A}_{W}^{(0)}$.

\begin{figure}[th]
\begin{center}
\epsfig{file=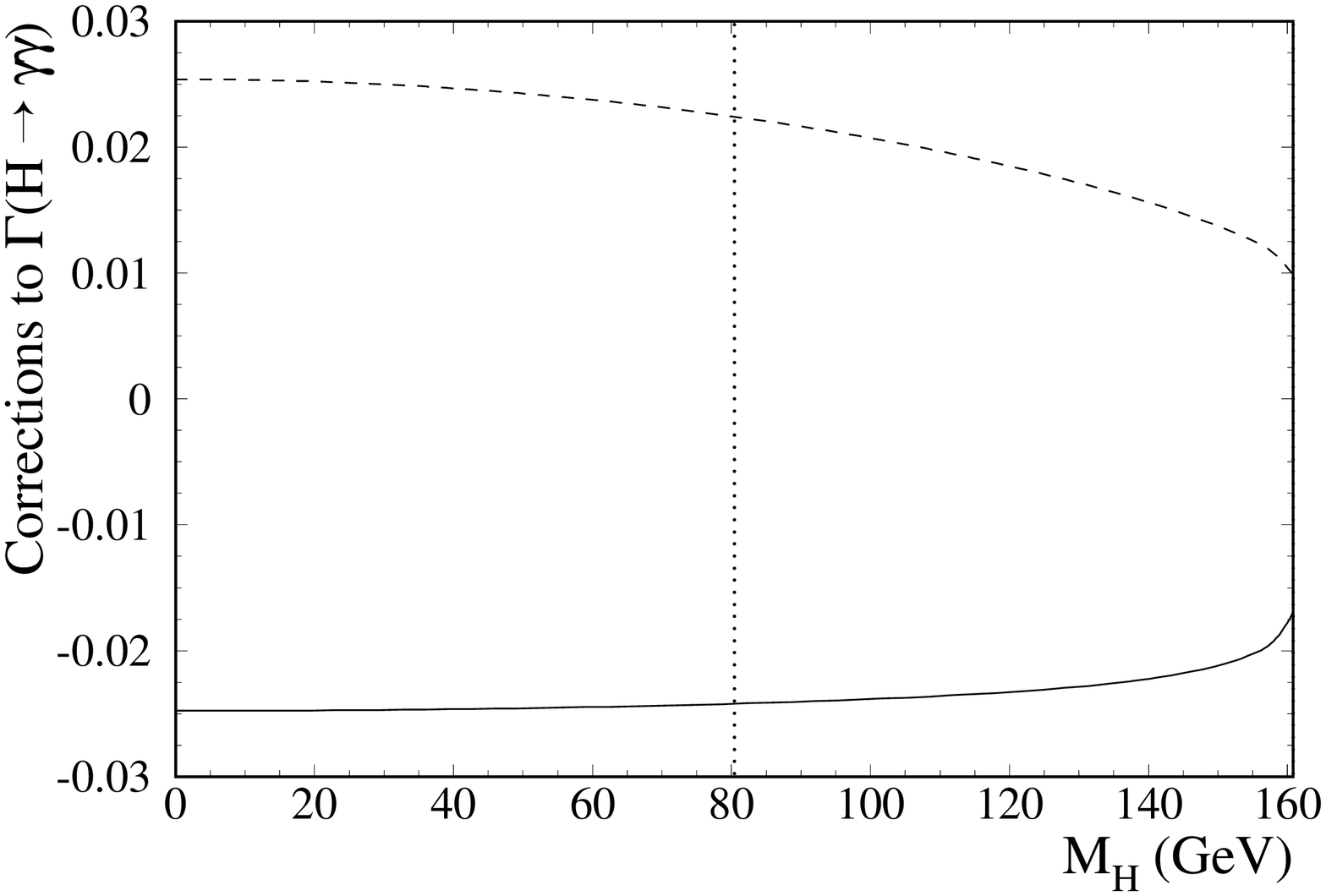,width=\textwidth}
\end{center}
\caption{\label{fig::LO_NLO} 
Dominant two-loop corrections to $\Gamma(H\to\gamma\gamma)$ as functions of
$M_H$.
The $\mathcal{O}(G_F M_t^2)$ electroweak correction (solid curve) is
compared with the $\mathcal{O}(\alpha_s)$ QCD one (dashed curve).
The dotted vertical line and the right edge of the frame encompass the
intermediate-mass range of the Higgs boson.}
\end{figure}

For the comparison with future measurements of $\Gamma(H\to\gamma\gamma)$, all
known corrections have to be included in Eq.~(\ref{Notation}).
In this connection, it is interesting to compare the size of the new
$\mathcal{O}(G_F M_t^2)$ electroweak correction with the well-known
$\mathcal{O}(\alpha_s)$ QCD one \cite{Djouadi:1990aj}.
This is done in Fig.~(\ref{fig::LO_NLO}), where the respective corrections to
$\Gamma(H\to\gamma\gamma)$ are displayed as functions of $M_H$.
As in Figs.~\ref{1loopW} and \ref{2loopT}, the dotted vertical line and the
right edge of the frame margin the intermediate-mass range of the Higgs boson.
We observe that, within the latter, the $\mathcal{O}(G_F M_t^2)$ correction
slightly exceeds the $\mathcal{O}(\alpha_s)$ one in magnitude, a rather
surprising finding.
Due to the sign difference, the two corrections practically compensate each 
other.
The two-loop electroweak correction induced by light-fermion loops, which has
become available recently \cite{Aglietti:2004nj}, is also negative, but has a
slightly smaller size than the $\mathcal{O}(G_F M_t^2)$ correction.


\section{\label{sec::summary}Conclusions}

We calculated the dominant two-loop electroweak correction, of
$\mathcal{O}(G_{F} M_{t}^{2})$, to the partial width of the decay into two
photons of the SM Higgs boson in the intermediate mass range,
$M_W\le M_H\le 2M_W$, where this process is of great phenomenological
relevance for the searches at hadron colliders of this elusive missing link of
the SM. 

We evaluated the relevant Feynman diagrams by application of the
asymptotic-expan\-sion technique exploiting the mass hierarchy
$2M_t\gg 2M_W\gg M_H$.
In this way, we obtained an expansion of the full
$\mathcal{O}(G_{F} M_{t}^{2})$ result in the mass ratio
$\tau_{W}=M_{H}^{2}/(2M_{W})^{2}$ through $\mathcal{O}(\tau_{W}^{4})$.

The convergence property of this expansion and our experience with the
analogue expansion at the Born level, where the exact result is available for 
reference, lead us to believe that these five terms should provide a very good
approximation to the exact result for $M_H\alt140$~GeV.
By the same token, the deviation of this approximation for the
$\mathcal{O}(G_{F} M_{t}^{2})$ amplitude $\mathcal{A}_{tW}^{(1)}$ from the
unknown exact result for this quantity is likely to range from 2\% to 20\% as
the value of $M_H$ runs from 140~GeV to $2M_W$.

In the intermediate Higgs-boson mass range, the $\mathcal{O}(G_{F} M_{t}^{2})$
electroweak correction reduces the size of $\Gamma(H\to\gamma\gamma)$ by
approximately 2.5\% and thus fully cancels the positive shift due to the
well-known $\mathcal{O}(\alpha_s)$ QCD correction \cite{Djouadi:1990aj}.

As a by-product of our analysis, we also recovered the
$\mathcal{O}(G_{F} M_{t}^{2})$ correction to the partial width of the decay
into two gluon jets of the intermediate-mass Higgs boson, in agreement with
the literature \cite{Djouadi:1994ge}.


\bigskip
\noindent
{\bf Acknowledgements}
\smallskip

We thank Paolo Gambino for a useful communication.
This work was supported in part 
by the Bundesministerium f\"ur Bildung und Forschung
through Grant No.\ 05~HT4GUA/4, 
by the Deutsche Forschungsgemeinschaft
through Grant No.\ KN~365/1-1,
by the Helmholtz-Gemeinschaft Deutscher Forschungszentren through
Grant No.\ VH-NG-008,
and by Sun Microsystems
through Academic Equipment Grant No.~EDUD-7832-000332-GER.


\end{document}